# Forensics in Industrial Control System: A Case Study

Pieter Van Vliet[2], M-T. Kechadi[1] and Nhien-An Le-Khac[1]

[1]School of Computer Science & Informatics, University College Dublin
Belfield, Dublin 4, Ireland
`{tahar.kechadi,an.lekhac}@ucd.ie`

[2], Rijkswaterstaat Security Centre, Ministry of Infrastructure and the Environment
Derde Werelddreef 1, Delft, The Netherlands
`pieter.van.vliet01@rws.nl`

**Abstract.** Industrial Control Systems (ICS) are used worldwide in critical infrastructures. An ICS system can be a single embedded system working stand-alone for controlling a simple process or ICS can also be a very complex Distributed Control System (DCS) connected to Supervisory Control And Data Acquisition (SCADA) system(s) in a nuclear power plant. Although ICS are widely used today, there are very little research on the forensic acquisition and analyze ICS's artefacts. In this paper we present a case study of forensics in ICS where we describe a method of safeguarding important volatile artefacts from an embedded industrial control system and several other sources.

**Keywords:** Industrial Control System, Forensic artefacts, forensic process, case study.

## 1      Introduction

Industrial Control Systems (ICS) are used worldwide in critical infrastructures. The term "critical infrastructure" conjures up images of highways, electrical grids, pipelines, government facilities and utilities. But the U.S. government definition also includes economic security and public health. The Department of Homeland Security defines critical infrastructure as "Systems and assets, whether physical or virtual, so vital to the United States that the incapacity or destruction of such systems and assets would have a debilitating impact on security, national economic security, national public health or safety, or any combination of those matters."[1].

In fact ICS is a broad term used in a variety of industries. ICS is not a single system but is a general term. As mentioned in the abstract, an ICS system can be a single embedded system like a Programmable Logic Controller (PLC) working stand-alone for controlling a simple process like an automatic door in an office building or an elevator in the same office building. ICS can also be a very complex Distributed Control System (DCS) connected to Supervisory Control And Data Acquisition (SCADA) system(s) [6] in a nuclear power plant.

On the other hand, there is little knowledge of ICS in the "forensic computer investigator world" resulting in a serious need for computer forensics to become more informed. Cybercrime related issues such as espionage and terrorism are challenges, but forensic investigations involving a fire in a chemical plant or explosion in electrical grid systems are increasing challenges for digital investigation units.

In most ICS systems a big part of the system are normal computers. Also the network diagrams and protocols are mostly just like normal ICT with conventional network setup and well-known protocols. For this part of the system traditional investigation methods are sufficient for digital investigation on the ICS system. Normal hard disk investigation, log analysis and network tools can be used to investigate what is running on the system and reveal the cause of the incident. However, standard forensics methodologies do not have inherent data collection capabilities for Programmable Logic Controller (PLC), Remote Terminal Unit (RTU), Intelligent Electronic Device (IED), or some other field-level device that is empowered to communicate through these communications mechanisms. In any ICS investigation, regardless of the age or uniqueness of the environment, collecting information from these field devices is a difficult task. Unlike normal ICT systems, ICS systems are designed with another perspective. ICS systems are designed with Safety in mind, rather than Security. The ICS timeline is also quite different in relation to ICT systems. An ICS system is designed for much longer periods of time. It is common for an ICS system to run for 20 or 30 years without update or upgrade.

With the experience of decades of forensic investigator career including many investigations related to ICS systems, authors discovered the lack of a framework for ICS forensics to safeguard the important information from the ICS system. Safeguard this information is a difficult task for computer forensics. PLC and DCS systems are embedded systems with their own operating systems and program languages. Dedicated hardware and many protocols are in use. Most digital forensic investigations techniques only cover conventional computer forensics and network investigations. Therefore, in this paper, we present an ICS Forensic process including important steps for acquiring important digital evidence for digital forensic investigation purposes. Not only the technical part of "how to safeguard" the important information is investigated, also the different jargon, perspective, timelines, goals and mind-set is described. We also focus on a case study of forensics in ICS system. The rest of this paper is organised as follows: Section 2 shows the related work of digital forensics for ICS systems. We discuss on ICS forensics challenges in Section 3, we also present an ICS process in this section. We describe and discuss a case study of ICS forensics in Section 4. Finally, we conclude and discuss on future work in Section 5.

## 2    Related work

Forensic ICS in literature are very often only related to SCADA systems. Some research topics contains more or less information about other devices like PLC, DCS, Industrial protocols (Modbus [7], Profibus [8] etc.) or security related issues on industrial control systems.

In 2013 Tina Wu et al [1] proposed SCADA Forensics Architecture and the increasing threat of sophisticated attacks on critical infrastructures, the limitations of using traditional forensic investigative processes and the challenges facing forensic investigators and there are no methodologies or data acquisition tools to extract data from embedded devices such as PLCs. There are data acquisition tools compatible with some field devices with the use of cables and flashing equipment, although this type of equipment is usually used for system servicing and repairs. This makes it difficult to obtain less common models of PLCs and RTUs and forensically sound access to the RAM and ROM on these devices is difficult to achieve without first turning the device off.

R. Barbosa [2] described Anomaly Detection in SCADA Systems, A Network Based Approach. He presented an extensive characterization of network traces collected in SCADA networks used in utility sector: water treatment and distribution facilities, and gas and electricity providers and note that despite the increasing number of publications in the area of SCADA networks, very little information is publicly available about real-world SCADA traffic. The number of attacks reported to the United States' Department of Homeland Security (DHS) grew from 9 in 2009, to 198 in 2011 and 171 in 2012.

R.M. van der Knijff [3] discussed on the difference between Control systems and SCADA forensics and the forensic knowledge and skills that are needed in the field of hardware, networks and data analysis. Assistance from experienced field engineers during forensic acquisition seems inevitable in order to guarantee process safety business continuity and examination efficiency. For specific control system components, there are currently no dedicated forensic tools to support acquisition and analysis of data. Instead of expensive and time-consuming physical examination methods, it is more practical to use existing tools and knowledge from the control system industry.

In 2008 U.S. Department of Homeland Security [4] provided a guidance for creating a cyber-forensics program for a control systems environment. This guidance described the challenges with collection, data analyses and reporting to industrial control systems. It identifies cautionary points that should be considered carefully when developing a cyber-forensics program for control systems. Many traditional device and control systems technologies do not provide for the collection of effective data that could be used for post-incident security analysis. An investigation harvesting evidence from core components that augment base operating systems should only be done with a full understanding of how the operating system has been changed. In addition, any auditing activity needs be carefully tested and deployed to assess any taxation on system resources. This guidance also describes what elements are important during investigations:

- Reference clock system
- Activity logs and transaction logs
- Other sources of data
- General system failures
- Real time forensics
- Device integrity monitoring
- Enhanced all-source logging and auditing

There is a project operating called the CRISALIS project [9]. CRISALIS aims at providing new means to secure critical infrastructure environments from targeted attacks, carried out by resourceful and motivated individuals. At the time of writing this paper the project is still ongoing and there is no working product available.

There are also open source tools for network security monitoring like the open-source Security Onion Linux suite [10], including Wireshark [11], NetworkMiner [12], etc. for network monitoring and intrusion detection. Passive network security monitoring is recommended as a key element to incident response. This will fit for the ICS environment because it is non-intrusive, so there is no risk of it disrupting critical processes or operations.

## 3   ICS Forensics

### 3.1   ICS Forensic Challenges

Forensic acquisition tools are widely available for conventional ICT systems like hard disks, volatile memory (RAM) and common consumer electronics like mobile phones and navigations systems. Similar tools do not exist for most ICS devices. Besides, in ICS systems, safety is the main goal rather than Security. If ICT people talk about Security and Safety in ICT systems they do mean:
- Firewalls to prevent hackers from entering the system since confidential information must be protected.
- Antivirus and Antimalware for protecting the users and the systems against viruses.
- Anti-spam to protect the users against spam in their mail.

However, If ICS people talk about Security and Safety in ICS systems they do mean:
- Protect the system against dangerous issues like wrong values in PLC's.
- Flow control and temperature sensors in the chemical plant.
- Voltage and current control in electrical grid installations.

Not only the other interpretation or different jargon can be an issue, also the difference between ICT people who are working mainly in the office or data centers and the ICS people working on the field inside the plant or control room. There is a big gap between the two different departments; other goals and other problems are creating completely different priorities on a daily base.

### 3.2   ICS Forensic Process

The purpose of our approach is to safeguard the important information from the ICS system. Depending if we talk about a running system which is still intact and connected to other devices like a distributed control system, or if we talk about a standalone control system like a single PLC or a post mortem investigation after a big incident like a fire or explosion in a chemical plant, several information sources are available to ac-quire important digital evidence for digital forensic investigation purposes. For this reason we have to setup an ICS Forensic process. Inside this process, we split up the information from two different sources:
- Network data
- Device data

**Network data acquisition**: For network data acquisition network investigation (depending on our investigation) we have to decide on what level (or levels) we need to analyze the network traffic.

Network Levels: A typical distributed ICS system has at least three different levels of network types:
- Device level such as sensor, programmable logic controller (PLC), actuator.
- Cell Level that is responsible to control the device controllers.
- Plant Level that is responsible to control the cell controllers.

Beside, network data can also be historical information like backup files, logging databases etc. Sources of network data can be listed as:
- Live Network Data (raw network data, arp tables, flow records, etc.)
- Historical Network Data (host based logs, database queries, firewall-logs,etc.)
- Other Log Files (backup archives, access point logs, historians, etc.)

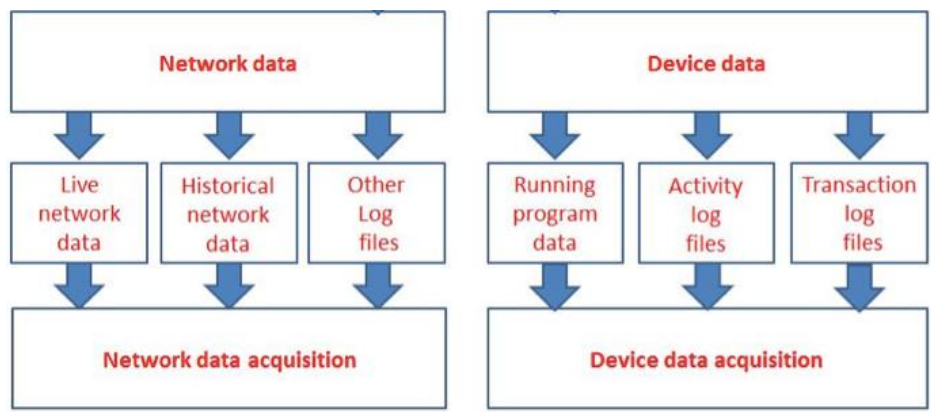

**Fig.1** ICS Forensic process

Not all tools or methods are safe to use in ICS environments. ICS systems often monitor or control processes in which a failure may have disastrous consequences (or may be otherwise very undesirable). For this reason active probing (like scanning for open ports and then opening arbitrary TCP connections) should generally be avoided.

**Device data acquisition:** Device data acquisition forensic tools do not exist for most ICS devices. Product specific service tools for programming a PLC, saving the program and servicing log files from a PLC to a service computer do exist. The question is can we use those service tools in a forensic matter to save important data from the PLC for later analyses? The sources of device data can be listed as:

- Running Program Data such as RAM dump, CHIP images, Memory cards…
- Activity Log Files such as RAM dump, active processes, control room logs, etc.
- Transaction Log Files such as Serial communication logs, Error logs, Event logs, etc.)

## 4      Case study

This case study relates to an ICS investigation of an incident in a Wind Turbine in October 2013. In this incident two service engineers died during a fire in the nacelle of a 70 m high wind turbine in a 12 wind turbines farm the Netherlands. During this investigation we found out how crucial some ICS systems related forensic investigations depend on volatile memory inside the PLC. During the fire all electronics from the nacelle were severely destroyed by the fire. The only device what was still intact was the ground controller inside the turbine tower on the ground level of the tower section. After the fire we removed the controller from the base of the turbine (Fig.2).

In close assistance with the Netherlands Forensic Institute, Department Digital Technology and the wind turbine manufacturer, we decided not to power-on the Cotas ground controller (Fig.3) because without all normal connected devices it would probably overwrite existing log entries with error messages due the missing devices. The wind turbine manufacturer later confirmed this. The battery pack attached to the ground controller contains two 1.5 volt batteries for power supply the battery operated RAM with 3 volt. During measurement we found out the battery power was dropped down to 2.2 volt.

At this stage it was not sure if this low voltage preserved all RAM content. Without opening the existing battery compartment, two extra battery pack holders ware connected in parallel with fresh batteries. This solution made it possible to replace the batteries one by one when needed if the voltage drops down again. To prevent data overwriting when powering up the ground controller again, the strategy was chosen in close assistance with the Netherlands Forensic Institute and the wind turbine manufacturer, to swap the PLC controller with a similar ground controller from a wind turbine in the same wind turbine farm as the destroyed one.

At this stage it was possible to make a copy of the RAM memory from the device. We used the service software and hardware from the wind turbine manufacturer. First we tested the software on another ground controller which was still in use in the same plant. After a successful test on this other ground controller we did the same on the ground controller from the destroyed turbine. The hardware was a serial to infrared converter to connect to the ground controller and setup a serial connection to the ground controller and download the configuration and log files from the ground controller. All configurations, alarm, system, production and warning logs including entries within the accident period were present so the 2.2 Volt had preserved all RAM content. The only line what did overwrite an entry in the log file was a normal start-up. During start up both ground controllers we did check the systems clock time with a real-time DCF clock to find out if there was a difference between the system time of the ground controller and the real time. For later analyses it was very important to know the exact time difference inside the log files and the real time (Table.1). A portable DCF [5] clock is an easy way to do this.

**Table 1.** Time difference measurements with a DCF-clock

| Ground Controller | Date-Time controller | Date-Tine DCF-lock | Difference |
|---|---|---|---|
| 15190 (turbine 12) | 12-12-2013 10:48:05 | 12-12-2013 10:23:47 | 00:24:18 |
| 15183 (turbine 2) | 12-12-2013 11:07:01 | 12-12-2013 10:42:42 | 00:24:19 |

After successfully download all configuration and log files from the ground controller we calculated a SHA-256 hash value from all downloaded files to verify that the saved data set has not been altered later during the investigation process.

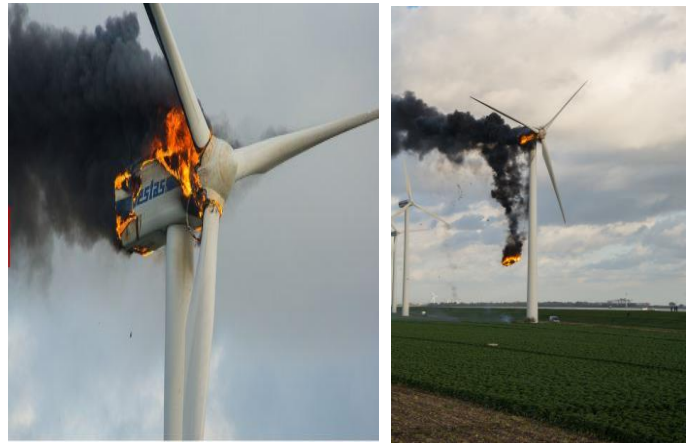

**Fig.2** Fire in Wind Turbine

Now we did successfully download all the important files from the ground controller it was possible to investigate the log files from this unit. The service software from the wind turbine manufacturer did have an option to save the files as .xml files and we were able read the content. The time and date inside this log files is not the real time, you have to convert all time lines with the difference measured during the start-up and the check we did with the DCF-clock.

Besides investigating the ground controller from the destroyed turbine, we also checked for other sources of digital evidence outside the turbine. One of the other sources was the connected SCADA server from the windmill park. Inside one of the other turbines there was a SCADA server installed in 2002. This server was a small industrial personal computer running on MS-DOS operating system. I did use a standard forensic method to make a forensic copy of the hard disk inside this SCADA server using a Tableau write-blocker and the forensic software FTK-imager. This SCADA server created reports of events, measurements and alarms. One of the events is shown below (Fig.4), this is an email from the SCADA server to the miller regarding an alarm.

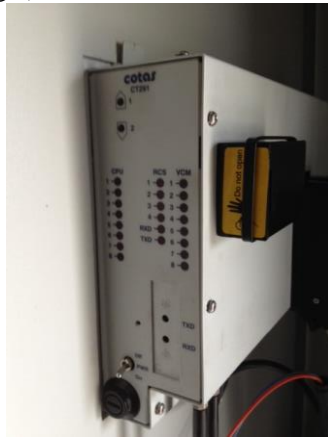 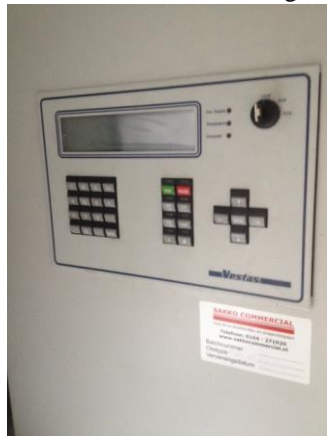

**(a)** Battery pack connected    **(b)** Front picture ground controller

**Fig.3** Cotas ground controller

One other source of digital evidence was the grid operator Stedin. The grid operator was able to deliver grid information about the time around the incident. The grid operator Stedin generated a report from the grid with a time sample of 10 seconds (Fig.5).

**Fig.4** Email from the SCADA server

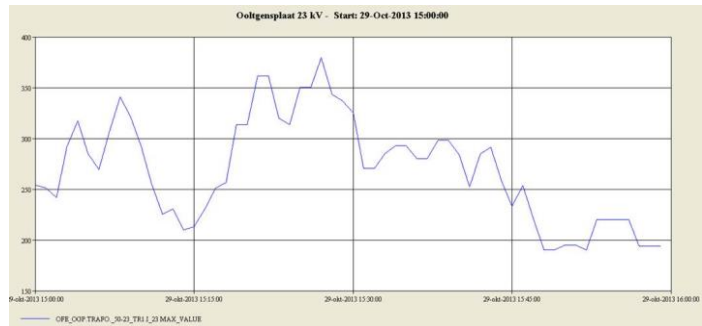

**Fig.5** Overview from grid operator

## 5   Conclusions and future work

Investigating ICS systems is not only about finding evidence of cybercrime related activities. Most investigations are incident related, like fire or explosions inside areas with ICS systems. Understanding the way how ICS systems are working, technical and tactical, is a must if you are involved with this kind of investigations. A big part of ICS systems is not different from normal conventional digital forensic work. Some parts of an ICS system are more difficult be course embedded systems like PLC's are different devices with their own communication protocols, connection interfaces, operating systems and program languages. Almost all manufactures of PLC's do have service tools that are able to safeguard RAM from the device, not always in a forensic certified way, but with some help of other tools you can get close to a forensically sound procedure in a manner that ensures it is "as originally discovered" and is reliable enough to be admitted into evidence.

In the future more research is desirable in reference systems and network flow. Anomaly detection [2] in ICS related networks is a very promising technique in securing ICS systems against cybercrime related crime. Especially ICS systems do have a static network flow and predictable behavior. Also in post mortem investigations like after an incident this predictable behavior can be a great help as long as you have a reliable source from the past. Anomaly detection technique can only work well if you know what is normal. Sources form the past can come from historians, pcap's, firewall log files, SCADA systems, etc. Also others sites with the same ICS system installed can be a good reference model for post mortem anomaly detection. Indeed, when dealing with the analysis of very forensics data extracted from ICS systems, data mining framework and knowledge map techniques [13][14][15] will be considered.

If you have a big and complex ICS system with remote sites and SCADA systems connected, it is not always necessary to investigate the whole system. Depending on your investigation you can make the system smaller and only investigate the PLC memory. Smaller do not always means simple, again a simple standalone PLC, without any connection to a SCADA system or any other device like a HMI, can still be a complex device. For example, if you don't know how to connect your tools to an embedded device like this, or if you are not familiar with the programming language in use. On the other hand, investigating the PLC RAM directly will give you more information about the PLC process. Most of the remote SCADA systems only record parts of the log files generated by the PLC. Some log files are not interesting for the process and will not be transmitted to the SCADA systems. In this case the only place where you can find all the needed information is inside the RAM on the PLC itself. Working together in close assistance with the manufacturer can help you to take the best actions and deliver the needed tools during your investigation.